\documentclass[preprint,12pt]{elsarticle}

\usepackage{epsfig}
\usepackage{here}
\usepackage{amssymb}

\usepackage{lineno}
\newcommand*\patchAmsMathEnvironmentForLineno[1]{
  \expandafter\let\csname old#1\expandafter\endcsname\csname #1\endcsname
  \expandafter\let\csname oldend#1\expandafter\endcsname\csname end#1\endcsname
  \renewenvironment{#1}
     {\linenomath\csname old#1\endcsname}
     {\csname oldend#1\endcsname\endlinenomath}}
\newcommand*\patchBothAmsMathEnvironmentsForLineno[1]{
  \patchAmsMathEnvironmentForLineno{#1}
  \patchAmsMathEnvironmentForLineno{#1*}}
\AtBeginDocument{
\patchBothAmsMathEnvironmentsForLineno{equation}
\patchBothAmsMathEnvironmentsForLineno{align}
\patchBothAmsMathEnvironmentsForLineno{flalign}
\patchBothAmsMathEnvironmentsForLineno{alignat}
\patchBothAmsMathEnvironmentsForLineno{gather}
\patchBothAmsMathEnvironmentsForLineno{multline}
}

\usepackage{color}
\usepackage{amsmath}

\journal{Physics Letters B}

\begin{document}

\begin{frontmatter}



\title{Search for neutrinoless quadruple beta decay of $^{136}$Xe in XMASS-I}

\address{\rm\normalsize XMASS Collaboration$^*$}
\author[ICRR,IPMU]{K.~Abe}
\author[ICRR,IPMU]{K.~Hiraide}
\author[ICRR,IPMU]{K.~Ichimura\fnref{KishimotoNow}}
\author[ICRR]{N.~Kato}
\author[ICRR,IPMU]{Y.~Kishimoto\fnref{KishimotoNow}}
\author[ICRR,IPMU]{K.~Kobayashi\fnref{KenkouNow}}
\author[ICRR]{M.~Kobayashi\fnref{MKobaNow}}
\author[ICRR,IPMU]{S.~Moriyama}
\author[ICRR,IPMU]{M.~Nakahata}
\author[ICRR]{K.~Sato}
\author[ICRR,IPMU]{H.~Sekiya}
\author[ICRR]{T.~Suzuki}
\author[ICRR,IPMU]{A.~Takeda}
\author[ICRR]{S.~Tasaka\fnref{TasakaNow}}
\author[ICRR,IPMU]{M.~Yamashita\fnref{YamashitaNow}}
\author[SNU]{B.~S.~Yang}
\author[IBS]{N.~Y.~Kim\fnref{NYKimNow}}
\author[IBS]{Y.~D.~Kim}
\author[IBS,KRISS]{Y.~H.~Kim}
\author[ISEE]{R.~Ishii}
\author[ISEE,KMI]{Y.~Itow}
\author[ISEE]{K.~Kanzawa}
\author[ISEE]{K.~Masuda}
\author[IPMU]{K.~Martens}
\author[IPMU]{A.~Mason\fnref{MasonNow}}
\author[IPMU]{Y.~Suzuki\fnref{SuzukiNow}}
\author[Kobe]{K.~Miuchi}
\author[Kobe,IPMU]{Y.~Takeuchi}
\author[KRISS]{K.~B.~Lee}
\author[KRISS]{M.~K.~Lee}
\author[Miyagi]{Y.~Fukuda}
\author[NH,IPMU]{H.~Ogawa}
\author[Tokai1]{K.~Nishijima}
\author[Tokushima]{K.~Fushimi}
\author[Tsinghua,IPMU]{B.~D.~Xu}
\author[YNU1]{S.~Nakamura}

\address[ICRR]{Kamioka Observatory, Institute for Cosmic Ray Research, the University of Tokyo, Higashi-Mozumi, Kamioka, Hida, Gifu, 506-1205, Japan}
\address[SNU]{Department of Physics and Astronomy, Seoul National University, Seoul, 08826, South Korea}
\address[IBS]{Center for Underground Physics, Institute for Basic Science, 70 Yuseong-daero 1689-gil, Yuseong-gu, Daejeon, 305-811, South Korea}
\address[ISEE]{Institute for Space-Earth Environmental Research, Nagoya University, Nagoya, Aichi 464-8601, Japan}
\address[IPMU]{Kavli Institute for the Physics and Mathematics of the Universe (WPI), the University of Tokyo, Kashiwa, Chiba, 277-8582, Japan}
\address[KMI]{Kobayashi-Maskawa Institute for the Origin of Particles and the Universe, Nagoya University, Furo-cho, Chikusa-ku, Nagoya, Aichi, 464-8602, Japan}
\address[Kobe]{Department of Physics, Kobe University, Kobe, Hyogo 657-8501, Japan}
\address[KRISS]{Korea Research Institute of Standards and Science, Daejeon 305-340, South Korea}
\address[Miyagi]{Department of Physics, Miyagi University of Education, Sendai, Miyagi 980-0845, Japan}
\address[NH]{Department of Physics, College of Science and Technology, Nihon University, Kanda, Chiyoda-ku, Tokyo, 101-8308, Japan}
\address[Tokai1]{Department of Physics, Tokai University, Hiratsuka, Kanagawa 259-1292, Japan}
\address[Tokushima]{Department of Physics, Tokushima University, 2-1 Minami Josanjimacho Tokushima city, Tokushima, 770-8506, Japan}
\address[Tsinghua]{Department of Engineering Physics, Tsinghua University, Haidian District, Beijing, China 100084}
\address[YNU1]{Department of Physics, Faculty of Engineering, Yokohama National University, Yokohama, Kanagawa 240-8501, Japan}

\cortext[CollabEMail]{{\it E-mail address:} xmass.publications17@km.icrr.u-tokyo.ac.jp}
\fntext[KishimotoNow]{Now at Research Center for Neutrino Science, Tohoku University, Sendai 980-8578, Japan}
\fntext[KenkouNow]{Now at Waseda Research Institute for Science and Engineering, Waseda University, 3-4-1 Okubo, Shinjuku, Tokyo 169-8555, Japan}
\fntext[MKobaNow]{Now at Institute for Space-Earth Environmental Research, Nagoya University, Nagoya, Aichi 464-8601, Japan}
\fntext[TasakaNow]{Now at Gifu University, Gifu 501-1193, Japan}
\fntext[YamashitaNow]{Now at Kavli Institute for the Physics and Mathematics of the Universe (WPI), the University of Tokyo, Kashiwa, Chiba, 277-8582, Japan}
\fntext[NYKimNow]{Now at Nuclear Research Institute for Future Technology and Policy, Seoul National University, Seoul, 08826, South Korea}
\fntext[MasonNow]{Now at Department of Physics, University of Oxford, Oxford, Oxfordshire, United Kingdom}
\fntext[SuzukiNow]{Now at Kamioka Observatory, Institute for Cosmic Ray Research, the University of Tokyo, Higashi-Mozumi, Kamioka, Hida, Gifu, 506-1205, Japan}

\begin{abstract}
A search for the neutrinoless quadruple beta decay of $^{136}$Xe was conducted with the liquid-xenon detector XMASS-I using $\rm 327\; kg \times 800.0 \; days$ of the exposure.
The pulse shape discrimination based on the scintillation decay time constant which distinguishes $\gamma$-rays including the signal and $\beta$-rays was used to enhance the search sensitivity.
No significant signal excess was observed from the energy spectrum fitting with precise background evaluation, and we set a lower limit of the half life of  3.7 $\times$ 10$^{24}$~years at 90\% confidence level.
This is the first experimental constraint of the neutrinoless quadruple beta decay of $^{136}$Xe.

\end{abstract}


\end{frontmatter}


%
%

\section{Introduction}
In spite of a great success of the standard model (SM) in particle physics,
the nature of the neutrino is not yet understood thoroughly.
If the neutrino is a Majorana particle, processes which violate the lepton number ($L$) by two ($\Delta$$L$ = 2) can take place.
The neutrinoless double beta decay (0$\nu\beta\beta$) is one of the $\Delta$$L$ = 2 processes beyond the SM.
The observation of the 0$\nu\beta\beta$ would tell us the Majorana nature of the neutrino. 
It is also linked to the seesaw mechanism, which explains the extremely light neutrino mass, and baryon number asymmetry in the universe via leptogenesis~\cite{Davidson:2008bu}.
Although a number of 0$\nu\beta\beta$ experiments using a variety of candidate nuclei have been performed, no evidence has been achieved so far~\cite{KamLAND-Zen:2022tow,PhysRevLett.123.161802,PhysRevLett.125.252502,PhysRevLett.124.122501,Arnold:2018tmo}.

On the other hand, Heeck and Rodejohann proposed that even if the neutrino is a Dirac particle, a decay which violates $L$ by four ($\Delta$$L$ = 4) can occur by adding three right-handed neutrinos in the SM~\cite{Heeck2013}. 
This $\Delta$$L$ = 4 process is naturally linked to the light Dirac mass terms of neutrinos~\cite{CHEN2013157}, CP violation~\cite{CHULIA2016431} and leptogenesis~\cite{PhysRevD.88.076004}.
The neutrinoless quadruple beta decay (0$\nu$4$\beta$)
\begin{equation} 
  \begin{aligned}
(A, Z) &\rightarrow (A, Z+4) + 4e^{-},
  \end{aligned}
\end{equation} 
is one of the $\Delta$$L$ = 4 processes. Here A and Z are the atomic mass number and atomic number, respectively. It is theoretically predicted that only three candidate nuclei, $^{150}$Nd, $^{136}$Xe, and $^{96}$Zr, can undergo this process.
The NEMO-3 experiment reported the result of a search for $^{150}$Nd 0$\nu$4$\beta$ decay~\cite{PhysRevLett.119.041801}, but there is no experimental search with either $^{136}$Xe or $^{96}$Zr 0$\nu$4$\beta$ decay so far.
The Q-value of the $^{136}$Xe 0$\nu$4$\beta$  ($Q_{0\nu 4\beta}$) is 79~keV~\cite{Wang_2017}\footnote[9]{The $Q_{0\nu4\beta}$ of $^{136}$Xe is mentioned as 44~keV in Ref.~\cite{Heeck2013}. The value is updated to 79.2 $\pm$ 0.4~keV in Ref.~\cite{Wang_2017}.}.
The XMASS-I accumulated more than 2 years' data at a low background (BG) rate of $O(10^{-4}$) counts/keV/day/kg in the energy region of interest~\cite{10.1093/ptep/pty053,20191,2018153} under a stable temperature and pressure of the liquid xenon (LXe)~\cite{PhysRevD.97.102006}.
These data are suitable for a $^{136}$Xe 0$\nu$4$\beta$ decay search with a high sensitivity.
In this paper, a first search for the 0$\nu$4$\beta$ decay in XMASS-I with a total exposure of $\rm 327\; kg\times 800.0 \; days$ is reported. 

\section{XMASS-I detector}
The XMASS-I detector is located in the Kamioka mine under 1,000~m of rock, corresponding to 2,700~meter water equivalent.
The inner detector (ID) contains 832~kg of LXe inside a pentakis-dodecahedron shape oxygen free high conductivity copper structure with an inscribed radius of about 43~cm.
Scintillation light from the LXe is detected by 630 2-inch hexagonal R10789 photomultiplier tubes (PMTs) and 12 2-inch cylindrical R10789Mod PMTs with a total photocathode coverage of 62.4\% of the detector's inner surface\footnote[10]{61.5\% with the consideration of the dead PMTs.}. 
Signals from the 642 ID PMTs were recorded by CAEN V1751 waveform digitizers with a sampling rate of 1~GHz. 
The radioactivity of the R10789 PMT is summarized in Ref.~\cite{ABE2019171}.
In order to reduce external $\gamma$-rays and neutrons from the surrounding rock, the ID is placed at the center of the outer detector (OD). The OD is a cylindrical tank 10~m in diameter and 11~m in height  filled with ultrapure water. 72 Hamamatsu 20-inch R3600 PMTs are mounted on the inner surface of the water tank to provide an active muon veto. 
More details of the detector can be found in Ref.~\cite{ABE201378}.

The individual ID PMT gains were continually monitored by blue LEDs embedded in the inner surface of the ID PMT holder. The LEDs are flashed by one-pulse-per-second signals from the global positioning system. 
Calibration data were taken every one or two weeks by inserting a $^{57}$Co source along the detector's $z$-axis to monitor the stability of the optical parameters described in Ref.~\cite{10.1093/ptep/pty053,20191, PhysRevD.97.102006,201945}.
The $\gamma$-ray and X-ray calibration sources of $^{57}$Co, $^{241}$Am, and $^{55}$Fe were used for the measurement of the scintillation decay time constant of electron recoils in LXe~\cite{takiya}, the evaluation of the scintillation efficiency of the detector, and energy calibration. 
Since the energy calibration was performed with $\gamma$-ray and X-ray sources and the visible energy for the same deposited energy depends on the particle, the electron-equivalent energy unit $\rm keV_{ee}$ is used in this analysis to represent the event energies. 
\section{Signal simulation}
In a neutrinoless quadruple beta decay of $^{136}$Xe, four $\beta$-rays with a total energy of
 79~keV are emitted simultaneously~\cite{Wang_2017}.
Since these $\beta$-rays are expected to deposit all their energy in the liquid xenon at each decay, a monochromatic peak is expected in the energy spectrum.
The event rate $R$ for a given half life of the decay ($T_{0\nu 4\beta}$) is calculated as
\begin{equation} 
  \begin{aligned}
R&=&\left(\frac{\ln 2 }{T_{0\nu 4\beta} }\right)\cdot \left( \frac{MN_a}{A}\right)\cdot k,
  \end{aligned}
\end{equation} 
where $M$ is the mass of the LXe in the fiducial volume (327~kg), $N_{a}$ is the Avogadro constant, 
$A$ is the atomic mass of $^{136}$Xe atom, and $k$ is the natural abundance of $^{136}$Xe (0.089).
The momentums of the four electrons from each $0\nu 4\beta$ decay were simulated by DECAY0 event generator~\cite{Tretyak}. 
In the DECAY0 event generator,
the four-dimensional energy distribution for a single electron $\rho_{1}$ can be found as
\begin{equation} 
\begin{split}
\rho_{1}(t_{1}) &= e_{1}p_{1}F(t_{1},Z)\int^{t_{0}-t_{1}}_{0}e_{2}p_{2}F(t_{2},Z)dt_{2}\\ &\times\int^{t_{0}-t_{1}-t_{2}}_{0}e_{3}p_{3}F(t_{3},Z)e_{4}p_{4}F(t_{4},Z)dt_{3},
\end{split}
\end{equation}
where $e_{i}$, $t_{i}$, and $p_{i}$ are the total energy, kinetic energy and momentum  of the $i$-th electron, respectively. $t_{0}$ is the total energy available in the 4$\beta$ process.  $F(t, Z)$ is the Fermi function~\cite{TretyakNEMOnote}.
The scintillation yield from the four electrons and the detector response were simulated by the XMASS Monte Carlo (MC) simulation based on Geant4~\cite{Agostinelli:2002hh}.
In the XMASS MC, the Doke model~\cite{doke} with a further correction based on the total energy deposition of the $\gamma$-ray calibration data  was used to take the energy-dependent scintillation photon yield into account~\cite{GammaScale}.

\begin{figure*}[htbp]
\begin{center}
\includegraphics[clip,width=6.7cm]{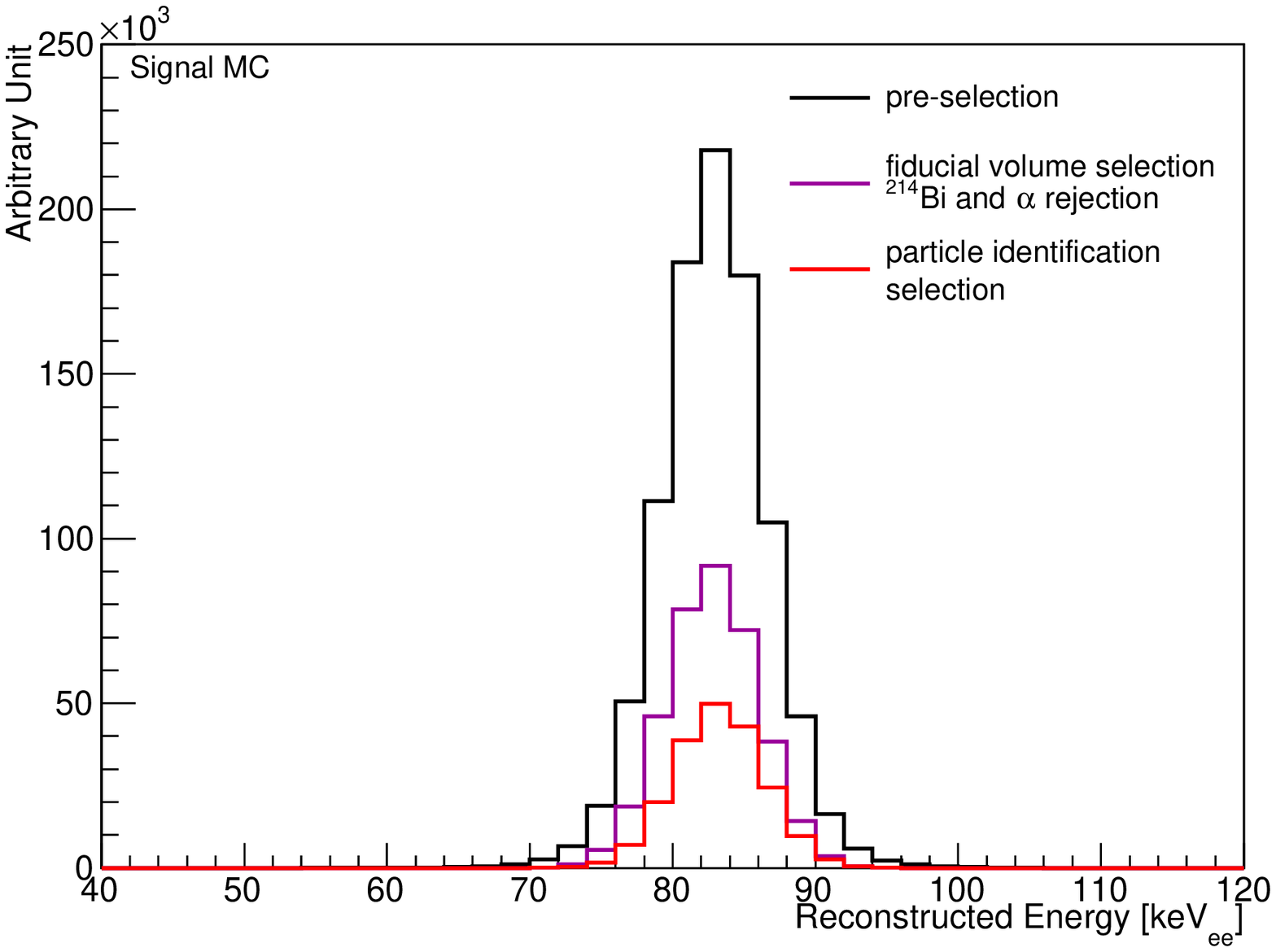}
\includegraphics[clip,width=6.7cm]{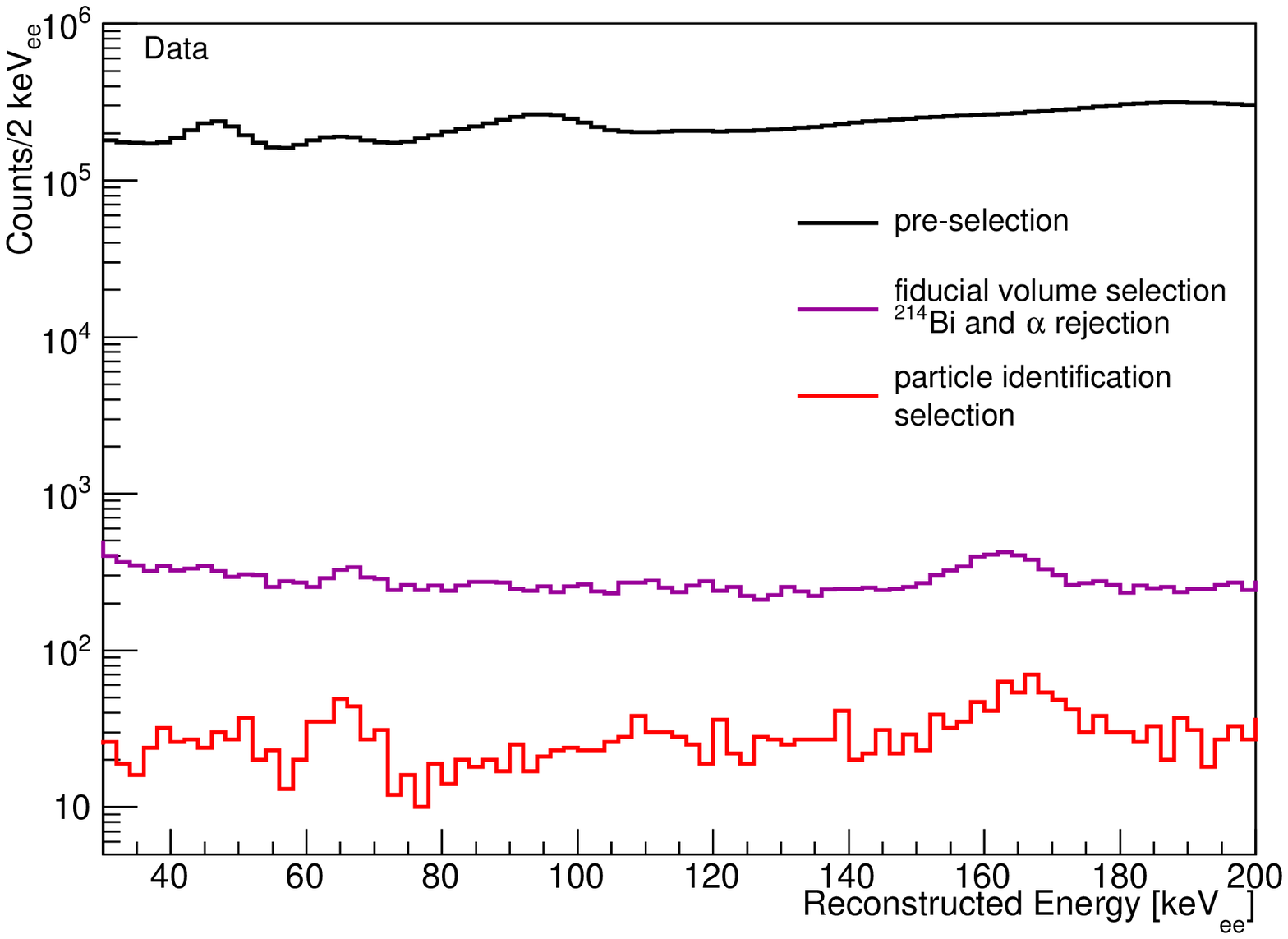}
\caption{ Energy spectra of the signal MC (left) and measured data (right) at various selection stages. Histograms after the pre-selection (black), the fiducial volume selection, $d\mathrm{T}_{\rm next}$ selection for removing $^{214}$Bi events, the scintillation time constant selection for removing $\alpha$ events (purple), and particle identification selection for selecting $\beta$-depleted sample ($\rm \beta CL<0.05$) (red) are shown. 
  The right-hand figure depicts the entire energy region used for the analysis 
($\rm 30 -  200\;  keV_{ee}$),
while the left-hand figure shows a magnified view of the $\rm 40 - 120\;  keV_{ee}$ region to make it easy to observe the 0$\nu$4$\beta$ signal. 
}
\label{classification}
\end{center}
\end{figure*}

Figure~\ref{classification} (left) shows the expected energy spectrum of the 0$\nu$4$\beta$ signal. The expected peak energy reconstructed from the observed light is seen at an energy 4\% higher than 79~keV. This is because the energy of each electron is lower than 79~keV. 
According to the Doke model~\cite{doke}, these low energy electrons are expected to have high light yields.
So the number of the observed total photoelectrons from four electrons with 79~keV total energy is larger than that of a single 79~keV electron.
The light yield of the 0$\nu$4$\beta$ signal has uncertainty on both of the increasing and decreasing sides.
It could decrease 3.0\% at maximum due to the scintillation model difference between the Doke model and the NEST model at a zero-electric-field~\cite{Szydagis_2013}.
It could increase 8.6\% at maximum due to the enhancement of the recombination probability~\cite{Szydagis_2011} since these four electrons are generated from one nucleus and their energies are deposited in a very small volume.
The uncertainty of the light yield of the 0$\nu$4$\beta$ signal is treated as a constrained parameter in the spectrum fitting described in Sec.~5.

\section{Data and event classification}
\label{refdataset}
A search for the $0\nu$4$\beta$ decay was carried out with 
the data accumulated from November 2013 to July 2016. The data set was same as that for our two-neutrino double electron capture on $^{124}$Xe and $^{126}$Xe~\cite{10.1093/ptep/pty053} analysis and the WIMP-$^{129}$Xe inelastic scattering search~\cite{20191}. 
During this period, the detector condition, specifically the temperature and pressure of LXe, was kept stable~\cite{PhysRevD.97.102006}. The data set was divided into four subsets (subset~1--4) depending on the Xe gas circulation status and the BG rate due to the neutron activation. 
The data within about 10 days after the neutron calibrations was not used since the BG rate due to the activation was high.
In the subset~1, the rate of the BG events from neutron-activated Xe isotopes was higher than other subsets because the data in this subset were taken only two weeks after the Xe installation and $^{252}$Cf neutron calibrations were performed twice.
The subset~2 started 60 days after the second $^{252}$Cf calibration. 
Neutron-activated peaks from $^{131\mathrm{m}}$Xe and $^{129\mathrm{m}}$Xe due to the $^{252}$Cf calibration disappeared and the rates of these peaks were stable in the subset~2.
Gas circulation for the Xe purification was started at the beginning of the subset~3. During the circulation, gas Xe was extracted from the detector, passed through a hot getter, condensed into liquid and returned to the detector.
The data in the subset~4 was taken after the removal of non-volatile impurities.
In this removal process, the Xe was recovered from the detector to the reservoir tank located outside the detector.
Then the evaporated xenon in the reservoir tank was returned to the detector through the hot getter.

The data went through several event selections.
First, the event is triggered only by ID. Then the event should have both the time difference from the previous event more than 10~ms and the root mean square of the hit timing in the event less than 100~ns. These selections are to remove the event caused by afterpulse following bright events.
For the rejection of external $\gamma$-rays, we apply the fiducial volume selection using the position reconstruction of the events' vertices.
In this reconstruction, measured light distributions on the PMTs were compared with the ones produced by the MC~\cite{ABE201378}.  
The events whose vertices were reconstructed inside the fiducial volume within 30~cm radius, were selected. 
This fiducial volume contains $\rm 327\; kg$ ($\rm 29\; kg$) of LXe ($^{136}$Xe).
The internal BG of $^{222}$Rn daughter nuclei were quantified by counting $^{214}$Bi events  taking the coincidences of $^{214}$Bi-$^{214}$Po chain-decays.
To take the coincidence, the time to the next event ($d\mathrm{T}_{\rm next}$) selection was used. By this coincidence, 99.6\% of the $^{214}$Bi events can be tagged by selecting events with $0.015\;{\rm ms}<d\mathrm{T}_{\rm next}< 1\; {\rm ms}$ since the half-life of $^{214}$Po is $164\;\mu{\rm s}$. 
The tagged events will be referred to as the $^{214}$Bi samples. In this sample, 0.4\% of non-$^{214}$Bi events were contaminated due to the accidental coincidence.

Further event selections were applied on non-tagged event to estimate the abundance of other BG and the signal. 
$\alpha$-rays are typical BG events observed in the non-tagged events.
A small fraction of scintillation light from $\alpha$ events would leak into the sensitive region via small gaps of the detector structure and would be detected by the PMTs. These events can be removed using the scintillation time constant obtained by the fitting of the summed-up waveforms with an exponential function. 
Those with scintillation time constants shorter than $\rm 30\; ns$ were classified as $\alpha$ events. This scintillation time selection eliminated almost all $\alpha$ events in the energy range above $30\; {\rm keV_{ee}}$, while more than 99.9\% of the signal events were retained. 

The events which survived the $\alpha$-event reduction were further classified into $\beta$-depleted and $\beta$-enriched samples using the particle identification selection. 
The particle identification uses the difference of the scintillation time profiles between $\beta$-rays and $\gamma$-rays.
The scintillation time constant of LXe has two components; $\tau_{s}$ = 2.2~ns, and $\tau_{t}$.
The smaller energy $\beta$-rays have shorter $\tau_{t}$ than the larger energy $\beta$-rays do~\cite{takiya}. 
$\gamma$-rays have shorter $\tau_{t}$ than $\beta$-rays with the same energy do since the $\gamma$-rays are converted into lower energy electrons via Compton and photoelectric effect.
For example, a 79~keV $\beta$-ray has $\tau_{t}$ of 38.5~ns. On the other hand, 79~keV $\gamma$-ray typically creates one 44~keV electron, one 25~keV electron, and a few~keV electrons by photoelectric effect.
44~keV and 25~keV $\beta$-rays have $\tau_{t}$ of 34.4~ns and 31.6~ns, respectively. These values are evaluated from Table~2 in Ref.~\cite{takiya}.
This classification makes the evaluation of the abundance of $\beta$-rays and $\gamma$-rays BGs easier by improving the significance of the $\gamma$-ray events.
Also the significance of the $0\nu 4\beta$ decay's signal is enhanced due to the shorter scintillation time of the signal 
since each electron has less energy than single $\beta$-ray events with the $Q_{0\nu 4\beta}$ energy. For example, 20~keV (a quarter of  the $Q_{0\nu 4\beta}$ energy) $\beta$-ray has $\tau_t$ of 30.3~ns. 

We introduce a particle identification parameter $\beta$CL defined as
\begin{equation}
\beta{\rm CL}=P\sum_{k=0}^{N-1}\frac{(-\ln P)^k}{k!}\;\;\;\; \left( P=\prod_{k=0}^{N-1}{\rm CDF}_\beta(E,t_k) \right)\; ,
\label{bcl}
\end{equation}
where $N$ is the number PMTs with pulses; $t_k$ is the timing of $k$-th pulse; $E$ is the event energy, and ${\rm CDF}_\beta (E,t)$ is the cumulative distribution function (CDF) for finding a pulse at time $t$ in a $\beta$-event of energy $E$ discussed in~\cite{10.1093/ptep/pty053}.
By definition, $\beta$CL distribution becomes uniform between 0 and 1 for $\beta$-ray events. On the other hand, a peak at 0 is expected for particles with shorter decay times than $\beta$-rays such as $\gamma$-ray and 0$\nu$4$\beta$ signal. 
$\alpha$ events also distribute around 0 in $\beta$CL distribution, although such events were already removed by the scintillation time selection.
Thus, the ratio of the $\gamma$-ray and the signal events to $\beta$-ray events
can be enhanced in the $\beta$-depleted sample.
The probabilities that $\beta$-ray, $\gamma$-ray including the 0$\nu$4$\beta$ signal are classified as $\beta$-depleted sample are referred to as $\beta$-ray mis-identification probability ($\beta$ mis-ID), and $\gamma$ acceptance, respectively. 
We set the threshold of the $\beta$CL ($\beta{\rm CL_{th}}$) at 0.05 for this analysis, following Ref.~\cite{10.1093/ptep/pty053}. 
The events with $\beta$CL less than 0.05 and greater than 0.05 are classified as $\beta$-depleted and $\beta$-enriched samples, respectively.
MC results showed that 53$\pm$16\% of 0$\nu$4$\beta$ signal events would remain after selecting events $\beta$CL less than 0.05, with 94$\pm$2\% of $\beta$-rays events from $^{214}$Bi events are rejected by this selection, corresponding to a signal-to-noise ratio improvement by a factor 9.
Here, the errors ($\pm$16\%, $\pm$2\%) correspond to the uncertainty of the $\beta$CL selection referred to as the $\gamma$ acceptance and the $\beta$ mis-ID in Sec.~5, respectively.
The signal MC and the data spectra after each treatment are shown in Figure~\ref{classification}. 

\begin{table}[htbp]
\begin{center}
\caption{Summary of the systematic uncertainty used as $p^{\rm const}_l$ in the spectrum fitting.}
  \begin{tabular}{lc}\hline 
    Item	& Fractional uncertainty  \\ 
    	& for each item  \\ \hline
$^{238}$U $\gamma$-rays BG from PMTs & $ \pm 9.4  \%$	     \\ 
$^{232}$Th $\gamma$-rays BG from PMTs & $ \pm 24  \%$	     \\ 
$^{60}$Co $\gamma$-rays BG from PMTs & $ \pm 11  \%$	     \\ 
 $^{40}$K $\gamma$-rays BG from PMTs & $ \pm 17  \%$	     \\ 
  $^{85}$Kr abundance in LXe & $ \pm 23  \%$	 \\ 
  Thermal neutron flux& $ \pm 27 \%$	     \\ 
Isotopic abundance of $^{136}$Xe & $\pm 1.3 \%$ \\
 Fiducial volume &   $\pm 4.5 \%$      \\ 
 Energy scale for $\beta$-depleted sample &   $\pm 2.0 \%$      \\ 
 Energy scale for $\beta$-enriched sample &   $\pm 2.0 \%$      \\
 $\gamma$ acceptance &  $ \pm 30 \%$  	     \\ 
Event increase due to dead PMT &  \\
   for 30 $\le$ E $\le$ 35 keV$_{\rm ee}$ & (7 $\pm$ 14 \%) \\
   for 35 $\le$ E $\le$ 40 keV$_{\rm ee}$ & (19 $\pm$ 16 \%) \\
 $\beta$ mis-ID &  Energy dependent \\
 & as shown in Fig.2  	\\
\hline 
  \end{tabular}
 \label{syserr} 
  \end{center}
\end{table}

\section{Energy spectrum fitting}
\label{secfit}
The events were classified into a $\beta$-depleted sample, a $\beta$-enriched sample, and a $^{214}$Bi sample in the previous section.
By fitting the energy spectra of these three samples simultaneously, we estimated the abundance of the signal and the BG.
To estimate the activities of the BG, the energy range of the fit was set from 30 to 200~$\rm keV_{ee}$. The energy bin's width is 2~$\rm keV_{ee}$. 
A $\chi^2$ value is defined as
\begin{equation}
\begin{split}
\chi^2=&-2\ln L \\
      =&2\sum_{i=1}^{N_{\rm sample}}\sum_{j=1}^{N_{\rm subset}}\sum_{k=1}^{N_{\rm bin}}
  \Biggl[ n^{\rm MC}_{ijk}(\{p^{\rm const}_l\},\{p^{\rm free}_m\})  \\
  & -n_{ijk}^{\rm data}+n^{\rm data}_{ijk}\ln \frac{n^{\rm MC}_{ijk}(\{p^{\rm const}_l\},\{p^{\rm free}_m\})}{n^{\rm data}_{ijk}} \Biggr] + \sum_{l=1}^{N_{\rm sys}}\frac{(1-p_l^{\rm const})^2}{\sigma _l ^2} \; ,
\end{split}
\label{chisquare}
\end{equation}
where $n_{ijk}^{\rm MC}$ is the  expected  number of events including the BG MC and signal MC, and $n_{ijk}^{\rm data}$ is the number observed of  events. 
The signal histogram was scaled by ${\tau^{-1}_{0\nu 4\beta}}$.
Indices ``$i$'', ``$j$'', and ``$k$'' mean $i$-th sample ($\beta$-depleted, $\beta$-enriched, and $^{214}$Bi), $j$-th subset, and $k$-th energy bin, respectively. 
Here, $N_{\rm sample}=3$, $N_{\rm subset}=4$, and $N_{\rm bin}=85$.
$p^{\rm const}_l\;(l=1,2,\cdots,N_{\rm sys} = 42)$ and $p^{\rm free}_m$ are constrained parameters and free parameters, respectively. 
The systematic uncertainty for $p^{\rm const}_l$ is $\sigma_l$.
The systematic uncertainties are summarized in Table~\ref{syserr}. 

The BGs in this study are the radioactive isotopes (RIs) in the PMTs ($^{40}$K, $^{60}$Co, $^{232}$Th and $^{238}$U), RIs uniformly distributed in the LXe ($^{214}$Pb, $^{214}$Bi, $^{136}$Xe 2$\nu\beta\beta$, $^{124}$Xe 2$\nu$ double electron capture, $^{85}$Kr, $^{39}$Ar and $^{14}$C) and RIs originating from thermal neutron captures ($^{133}$Xe, $^{131\mathrm{m}}$Xe and $^{125}$I). 
The constraints for RIs in the PMTs and $^{85}$Kr were determined based on the results of the BG study in XMASS~\cite{201945}.
The constraints for the abundance of  $^{214}$Pb and $^{214}$Bi, $^{124}$Xe 2$\nu$ double electron capture, $^{39}$Ar, and $^{14}$C were not given and their abundances were determined by the fitting.
Here, $^{39}$Ar is thought to be in argon gas which was used for a leakage test of the detector before starting data taking, and to have been absorbed in the detector material~\cite{10.1093/ptep/pty053}.
To account for the $^{133}$Xe and $^{131\mathrm{m}}$Xe BG amount related to the special work of neutron calibration and purification work, we introduce the free parameters of the abundance of $^{133}$Xe and $^{131\mathrm{m}}$Xe in the fitting.
The 2$\nu$2$\beta$ BG events of $^{136}$Xe were constrained by the result of KamLAND-Zen experiment~\cite{PhysRevLett.122.192501}. 
Thermal neutron flux measurements~\cite{otani,minamino} gave constraints of neutron-induced RI of $^{125}$I.
Since detector condition affects the amount of $^{85}$Kr and BG from thermal neutron flux, we considered the subset dependence of these two BGs.

\begin{figure}[htbp]
\begin{center}
\includegraphics[clip,width=7.5cm]{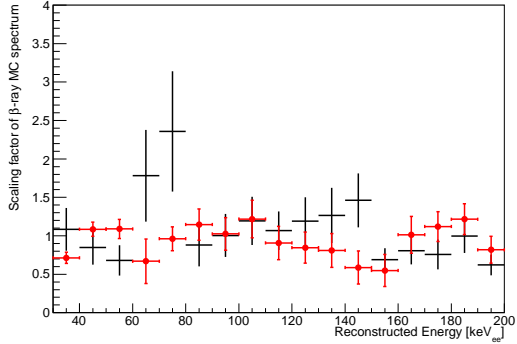}
\caption{Scaling factor of $\beta$-ray MC for the correction of $\beta$ mis-ID events. 
Black points and their error bars represent the means and errors calculated from the difference between MC and the data. 
In the fitting process, $\beta$-ray MC spectra were scaled by factors ``${\rm (mean)}+(p_l^{\rm const}-1)({\rm error})$'' depending on the energy. The red points are the scale factor for the best fit and 1$\sigma$ error bars.
}
\label{betaMisID}
\end{center}
\end{figure}

The uncertainty of the isotopic abundance of $^{136}$Xe was evaluated from a modified VG5400/MS-III mass spectrometer measurement at the Geochemical Research Center, the University of Tokyo~\cite{BajoMS}. The abundance was consistent with that of the natural xenon in the air.
The uncertainty of $\pm$1.3\% in the measurement was treated as the systematic uncertainty of the abundance of $^{136}$Xe isotope.
The comparison of the MC and the data of $\gamma$-rays from $\rm ^{241}Am$'s 59.5~keV $\gamma$-rays and $\rm ^{57}Co$'s~122 keV
$\gamma$-rays was used for the evaluation of the uncertainty of the fiducial volume ($\pm$4.5\%), the energy scale for $\beta$-depleted and $\beta$-enriched sample (both $\pm$2\%), and $\gamma$ acceptance  ($\pm$30\%)~\cite{10.1093/ptep/pty053,20191}.
Since energy scale varies depending on the detector condition, we introduced the energy scale parameter in each subset in Eq.~\ref{chisquare}. 
The dead PMT effect (7\% $\pm$ 14\% for  30 $<$ E $<$ 35~keV$_{\rm ee}$ and 19\% $\pm$ 16\% for 35 $<$ E $<$ 40~keV$_{\rm ee}$ region), accounts for the mis-reconstructed events increase due to dead PMTs discussed in Refs.~\cite{2018153,201945,2020135741}. 
The uncertainty of the $\beta$ mis-ID was evaluated by the comparison of the data and MC of $^{214}$Bi events selected by the  $^{214}$Bi--$^{214}$Po coincidence.
Using the $\beta$-ray's continuous spectrum over the energy region of this analysis, the energy-dependent uncertainties were obtained.
The energy region from 30 to 200~$\rm keV_{ee}$ was divided into 17 bins. Probabilities of $\beta$ mis-IDs for data and MC of $^{214}$Bi delayed coincidence events were compared in each bin. 
To correct the difference of $\beta$ mis-ID between the data and MC, $\beta$-ray BG MC histograms were scaled energy-dependently in the fitting using the $\beta$ mis-ID ratio obtained from $^{214}$Bi events between the data and MC shown in Fig.~\ref{betaMisID}. 
The uncertainty of the light yield of the 0$\nu$4$\beta$ mentioned in Sec.~2 (-3.0\%, +8.6\%) is treated as a constrained parameter without the penalty $\chi^{2}$ term in Eq.~\ref{chisquare}.

\section{Results}
The energy spectra of three samples ($\beta$-depleted, $\beta$-enriched, and $^{214}$Bi samples) were fitted with the MC spectra of the signal and BG simultaneously so that the signal abundance and the BG abundances were determined at the same time.
No significant signal excess over the expected BG was found and the null 0$\nu$4$\beta$ signal case gave the best-fit with $\chi^{2}$/ndf = 1096/997.
Thus, the 90\% confidence level (CL) lower limit on the $T_{0\nu 4\beta}$ was calculated by the following equation.

\begin{equation}
  \frac{\int_0^{{T_{0\nu 4\beta,90}}^{-1}}L({T_{0\nu 4\beta}}^{-1})\; d{T_{0\nu 4\beta}}^{-1} }{\int_0^{\infty}L({T_{0\nu 4\beta}}^{-1})\; d{T_{0\nu 4\beta}}^{-1}}=0.9
\end{equation}

\begin{figure*}[htbp]
\begin{center}
\includegraphics[clip,width=15.0cm]{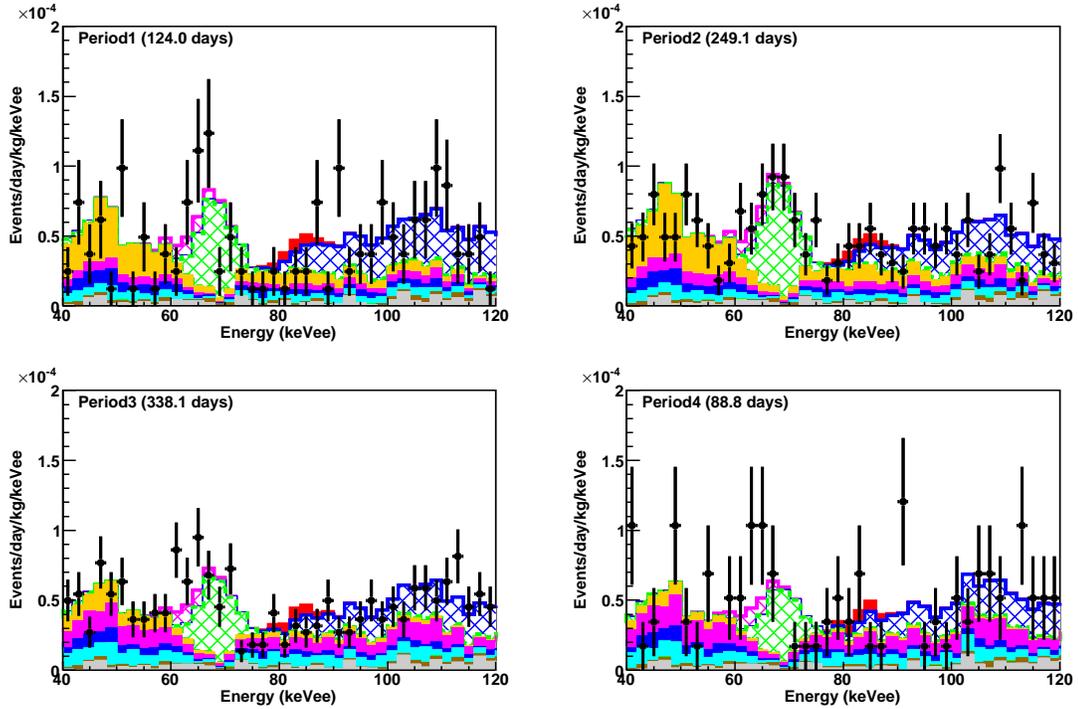}
\caption{Energy spectra for $\beta$-depleted samples for the four subsets. The data are shown as black points with statistical error bars. Signal (90\% CL lower limit of the half life, red filled), $^{126}$Xe two neutrino double electron capture (magenta hatched), $^{125}$I (green hatched), $^{133}$Xe (blue hatched) $^{14}$C (orange filled), $^{39}$Ar (magenta filled), $^{85}$Kr (blue filled), $^{214}$Pb (cyan filled), $^{136}$Xe 2$\nu$ double beta (brown filled), and external $\gamma$-rays (gray filled) backgrounds are shown as stacking histograms. Here, we show an enlarged view of the signal region for the convenience.}
\label{zoom90}
\end{center}
\end{figure*}

The calculated 90\% CL lower limit is 3.7 $\times$ 10$^{24}$ years.
Measured $\beta$-depleted samples' spectra (black histograms) and best-fit BG spectra (color histograms) are shown in Figure~\ref{zoom90}.
It should be noted that 0$\nu$4$\beta$ spectra  with the size of the 90\% CL lower limit of the half life (red) are added on top of the best-fit BG spectra to illustrate the contribution to the BG ones.
The LXe purification started at the beginning of subset~3 decreased the activity of $^{14}$C (orange) as time proceeds. The cause of the increasing of $^{39}$Ar (magenta) was presumably due to the emanation from the inner structure of the detector~\cite{10.1093/ptep/pty053}.
This is the first experimental constraint on the half life of ${0\nu 4\beta}$ decay of $^{136}$Xe and the longest half life limit of the ${0\nu 4\beta}$ decay.

\section{Conclusion}
A search for the neutrinoless quadruple beta decay of $^{136}$Xe was conducted using $\rm 327\; kg\times 800.0\; days$ of XMASS-I data.
The particle identification based on the difference of the scintillation time constant between $\gamma$-rays and $\beta$-rays enhanced the sensitivity.
The particle identification parameter ($\beta$CL) and timing-coincidence ($d\mathrm{T}_{\rm next}$) was used to create three separate subsets of the data ($\beta$-enriched, $\beta$-depleted, and $^{214}$Bi sample). These spectra were fit with the BG MC + 0$\nu$4$\beta$ signal MC.
In this analysis, no significant excess over the expected BG was found and a 90\% CL lower limit of the half life of 3.7 $\times$ 10$^{24}$~years for the neutrinoless quadruple beta decay of $^{136}$Xe was set.
This is the first experimental constraint on the neutrinoless quadruple beta decay of $^{136}$Xe.

\section*{Acknowledgements}
We gratefully acknowledge the cooperation of the Kamioka Mining and Smelting Company.
This work was supported by the Japanese Ministry of Education,
Culture, Sports, Science and Technology, Grant-in-Aid for Scientific Research, 
JSPS KAKENHI Grant No. 18K03669, 19GS0204, 26104004, and 19H05805,
the joint research program of the Institute for Cosmic Ray Research (ICRR), the University of Tokyo,
and partially by the National Research Foundation of Korea Grant funded
by the Korean Government (NRF-2011-220-C00006), and the Brain Korea 21 FOUR Project grant funded by the Korean Ministry of Education.

\bibliographystyle{elsarticle-num2}
\bibliography{Reference}

\end{document}